\documentstyle[11pt,paspconf,epsf]{article}

\begin{document}
\title{Distances and absolute magnitudes from 
trigonometric parallaxes}
\author{Fr\'ed\'eric Arenou}
\affil{DASGAL/URA 335 du C.N.R.S. -- Observatoire de Paris-Meudon,
	F-92195 Meudon Cedex, France -- E-mail: Frederic.Arenou@obspm.fr}
\author{Xavier Luri}
\affil{Departament d'Astronomia i Meteorologia, 
	Universitat de Barcelona, Avda. Diagonal
	647, E-08028, Barcelona, Spain -- E-mail: xluri@am.ub.es}
%

\begin{abstract}
In astrophysical applications, derived quantities like distan\-ces,
absolute magnitudes and velocities are used instead of the observed
quantities, such as parallaxes and proper motions. As the observed values
are affected by random errors and selection effects, the estimates of the
astrophysical quantities can be biased if a correct statistical treatment
is not used. This paper presents and discusses different approaches to
this problem. 

We first review the current knowledge of Hipparcos systematic and random
errors, in particular small-scale correlations.  Then, assuming Gaussian
parallax errors and using examples from the recent Hipparcos literature,
we show how random errors may be misinterpreted as systematic errors, or
transformed into systematic errors. 

Finally we summarise how to get unbiased estimates of absolute magnitudes
and distances, using either Bayesian or non-parametrical methods.  These
methods may be applied to get either mean quantities or individual
estimates.  In particular, we underline the notion of astrometry-based
luminosity, which avoids the truncation biases and allows a full use of
Hipparcos samples. 
\end{abstract}


%
\newcommand{\piH}{\pi_{\mathrm H}}
\newcommand{\rH}{r_{\mathrm H}}
\newcommand{\MH}{M_{\mathrm H}}
\newcommand{\sigH}{\sigma_{\piH}}
\newcommand{\piP}{\pi_{\mathrm P}}
\def\Lynga{Lyng{\aa}}
\def\spipi{{\sigma\over\pi}}
\def\spipiH{{\sigma\over\piH}}
\def\obs{{\mathrm O}}
\def\apriori{\it a priori}{}
\def\aposteriori{{\it a posteriori}}
\def\etal{et al.}
%

%
\section{Introduction}
%
Many papers have been devoted along the years to the various biases that
can arise in the determination of stellar luminosities from trigonometric
parallaxes.  The advent of the Hipparcos Catalogue, with its unprecented
accuracy and homogeneous data, could have been the occasion to efficiently
take these biases into account.

It seems, on the contrary, that in the majority of recent papers the
sample selections have been mostly based on the parallax relative
precision (based on ${\sigma\over{\piH}}$, where $\piH$ denotes the
Hipparcos parallax and $\sigma$ its formal precision) while it is well
known that sample truncations on the parallax relative error lead to
biased estimates of quantities derived from the parallax. Furthermore, the
various adopted limits on ${\sigma\over{\piH}}$ are merely a balance
between the expected precision on the resulting absolute magnitude and the
size of the sample and thus are not based on any statistical criteria.  Some
illustrative examples are shown in section \ref{literature}

The effects of random errors will be thoroughly discussed in the following
sections, but it is interesting to summarise here what a truncation on the
``observed'' relative error $\sigma\over\piH$ implicitly implies for the
resulting sample:
\begin{itemize}
 \item the truncation on $1\over\piH$ should produce an approximate 
  volume-limited sample, but the error on $1\over\piH$ is correlated with 
  the error on the absolute magnitude, implying a bias on this quantity; 
 \item for a given $1\over\piH$, the precision $\sigma$ is mainly due to
 photon noise, so that brighter stars will be preferentially selected;
 \item for a given apparent magnitude $\sigma$ also depends on the ecliptic
 latitude (due to the Hipparcos scanning law) thus adding a spatial
 selection;
 \item the values of $\sigma$ used are not the ``real'' values of
 the precision but its estimates. Thus, $\sigma\over\piH$ is the
 combination of two random variables, making the truncation more
 statistically complex; 
 \item it should also be taken into account that the initial sample
 (before truncation) represents  the content of the Hipparcos Catalogue.
 Apart from the ``Survey'', which is rather well defined, the other
 selections from which the Catalogue was built are not always clear
 (e.g.  a kinematical bias for nearby stars).  Further observing problems
 may have also happened, leading to the rejection of other stars.
\end{itemize}
\indent The final effect is that there is in fact {\it no} knowledge of the
representativeness of the sample with respect to the parent population.
Any statistics computed from this sample using parallaxes will probably be
a biased estimate of the quantity which one would like to obtain for the
parent population. Furthermore, if a generic (in the sense of not
specifically adapted to the characteristics of the sample) {\aposteriori}
bias correction is applied, the accuracy of the result is hardly
predictable. This is e.g.  the case for the Lutz-Kelker (1973) correction,
which assumes an uniform stellar density, whereas this assumption may not
be realistic.  Even if it is adequate, the confidence interval of the
correction may be very large (Koen, 1992), so that the precision on
absolute magnitude will be rather poor.

In general, truncating in parallax relative error in the hope of
benefiting from smaller random errors finally gives greater systematic
errors.  Moreover, the rejection of stars with high relative errors wastes
a large amount of data, from which the random errors could have been
reduced. Anticipating the conclusion of this paper, it must be noted
that no selection on the observed parallaxes should be done.  It should
also be remembered that the observing list of the brighter stars in the
Hipparcos Catalogue (Survey) was defined on purpose, in order to benefit
from clearly defined samples.  When it applies, the selection on apparent
magnitude may then be taken into account in the estimation procedure.  The
effect of this selection (Malmquist bias) is discussed in
Section~\ref{Malmquist}

%
\section{The error law of Hipparcos parallaxes}
%
Since the effect of random parallax errors on derived quantities will be
discussed in section~\ref{effects}, we review in this section the general
properties of Hipparcos errors.

%
\subsection {Gaussian errors}
%
It has been shown in various papers (e.g.  Arenou {\etal}, 1995, 1997) 
that in general the random errors of the Hipparcos parallaxes may be
considered Gaussian.  This may be seen for instance when these parallaxes
are compared to ground-based values of similar precision, or to distant
stars using photometric estimates, using the normalised differences
(parallax differences divided by the square root of the quadratic sum of
the formal errors).  This nice property of the parallax errors may then
fully be used in parametrical approaches which make use of the conditional
law of the observed parallaxes given the true parallax. 

The particular case of systematic errors at small angular scale will be
discussed in section \ref{smallscale}; for an all-sky sample, one may safely
consider that the global systematic error is small ($\leq 0.1$mas), that
the formal errors are good estimates of the random error dispersions and
that the random errors are uncorrelated from star to star . 

%
\subsection {Non-Gaussian errors}
%
Due to their Gaussian behaviour, random errors in the Hipparcos Catalogue
are of course expected to produce a number of stars whose astrometric
parameters deviate significantly from the $1\sigma$ error level so that,
of course, some hundredths of stars are expected to have an observed
parallax which deviates, say, 3 mas from the true parallax value. This is
a logical consequence in a large Catalogue like Hipparcos. 

In a few cases, however, it may happen that the error on the Hipparcos
data is much higher than expected. Although these are probably rare cases,
they have been mentioned for the sake of completeness in the Hipparcos 
documentation and illustrated here. 

Apart from the Double and Multiple Star Annex (DMSA, see ESA 1997), most
of the Hipparcos Catalogue is constituted by stars assumed to be single.
One obvious source of outliers may thus be undetected short period
binarity, since in this case the astrometric path of the star will not
exactly follow the assumed single star model (5 parameters:  position,
parallax, linear proper motion).

Two extreme cases of astrometric binaries are discussed below, which may
have been biased respectively in parallax and proper motion.  It must be
stressed that these cases are statistically rare and chosen for the
purpose of illustration, and that the duplicity had in fact been detected
by Hipparcos and flagged in the Catalogue.

The first example concerns HIP 21433, one of the 1561 Hipparcos stochastic
solutions (DMSA/X), where an excess scatter of the measurements may be
interpreted as the signature of an unknown orbital motion.  Indeed, this
star is a spectroscopic binary.  The interesting fact is that the period
is 330 days, i.e.  close to one year, so that there may have been a
confusion between the parallactic and orbital motion.  Adopting the 4
known orbital elements ($P$,$T$,$e$,$\omega_1$)  from Tokovinin {\etal}
(1994), the intermediate astrometric data has been re-reduced taking into
account the 5 astrometric parameters and the 7 orbital parameters, and the
new parallax found is $30.36\pm 0.87$ mas, instead of $34.23\pm 1.45$ mas
as in the published stochastic solution.  The parallax from the Hipparcos
solution, which does not take into account the binarity, has thus possibly
been biased, due to the $\approx 1$ year orbital motion. 

The second example concerns HIP 13081, one of the 2622 acceleration
solutions (DMSA/G), where the motion has not been linear during the
mission, interpreted as a binary of longer period.  When accounting for
the orbital motion, using data from Tokovinin (1992), a new solution has
been computed.  The inclination angle $i$ is near $90\deg$ so that the
path on the sky is nearly linear.  The proper motion of the barycentre is
$275\pm 3$ mas/yr instead of the published solution, $264\pm 1$ mas/yr. 
Strictly speaking, this is not a bias, since Hipparcos measured the
photocentre of the system, not the barycentre. 

In both examples, compared to the ``true'' value, the published solution
is significantly different from what could be expected in the case of
Gaussian errors.  Although these examples are unfavourable cases, it must
be pointed out that they were detected during the Hipparcos data
reduction.  The same effects may also be present for some other stars
where the binarity has not yet been detected, but this implies at the same
time that the astrometric perturbation is smaller.

%
\subsection{Small-scale systematic errors}\label{smallscale}
%
The operation mode of the Hipparcos satellite implied that the stars
within a given small field were frequently observed together with the same
complementary set of stars in the other field of view.  This introduces
correlations between the astrometric parameters of stars within some
square degrees but, due to the rather low sky density of Hipparcos, it is
not a problem, except for open star clusters.  This effect was studied
before the satellite launch by Lindegren (1988) and confirmed using the
final results by Lindegren {\etal} (1997) and Arenou (1997).

A special data reduction process had then to be used for cluster stars. 
This has been done in van Leeuwen (1997a,b) and Robichon {\etal} (1997),
and for this purpose the angular correlations have been calibrated, as
detailed in van Leeuwen \& Evans (1998) and Robichon {\etal} (1999). 

Although the correlation effect was known and taken into account, it was
possibly not realized that, for a single realisation of a given cluster,
this could mean a systematic error for the 
individual cluster members.  It must however
be remembered that the Hipparcos data was reduced by two different
Consortia, and the systematic error is probably not the same for
both, so that the merging of the two sets (Arenou, 1997) probably
reduced the effect.

In order to illustrate this correlation, one may take the extreme example
of NGC 6231, where all 6 Hipparcos stars have a negative parallax, whereas
the photometric estimate is $0.71\pm 0.02$ mas (Dambis, 1998).  A straight
weighted average of individual parallaxes would give $-0.71\pm 0.39$ mas;
even taking into account the correlations, the mean cluster parallax is
$-0.62\pm 0.48$ mas, which is still significantly different from the
photometric estimate. 

Apart from this extreme example, the question is whether the correlation
effect has correctly been accounted for in the estimation of the Hipparcos
mean parallax of each cluster.  Although this is not easy to prove for one
given cluster, an indirect statistical evidence may be obtained using a
sample of distant clusters. To carry out this test, clusters farther than
300 pc with at least 2 Hipparcos members were used: 66 clusters in Dambis
(1998)  Catalogue and 102 clusters in Loktin \& Matkin (1994) were found.
Concerning the latter, a 5\% parallax relative error has been adopted. The
mean photometric parallax error of these samples is about 0.04 mas, so
that the comparison with Hipparcos parallaxes shows mainly the Hipparcos
errors.

\begin{figure}
\plottwo{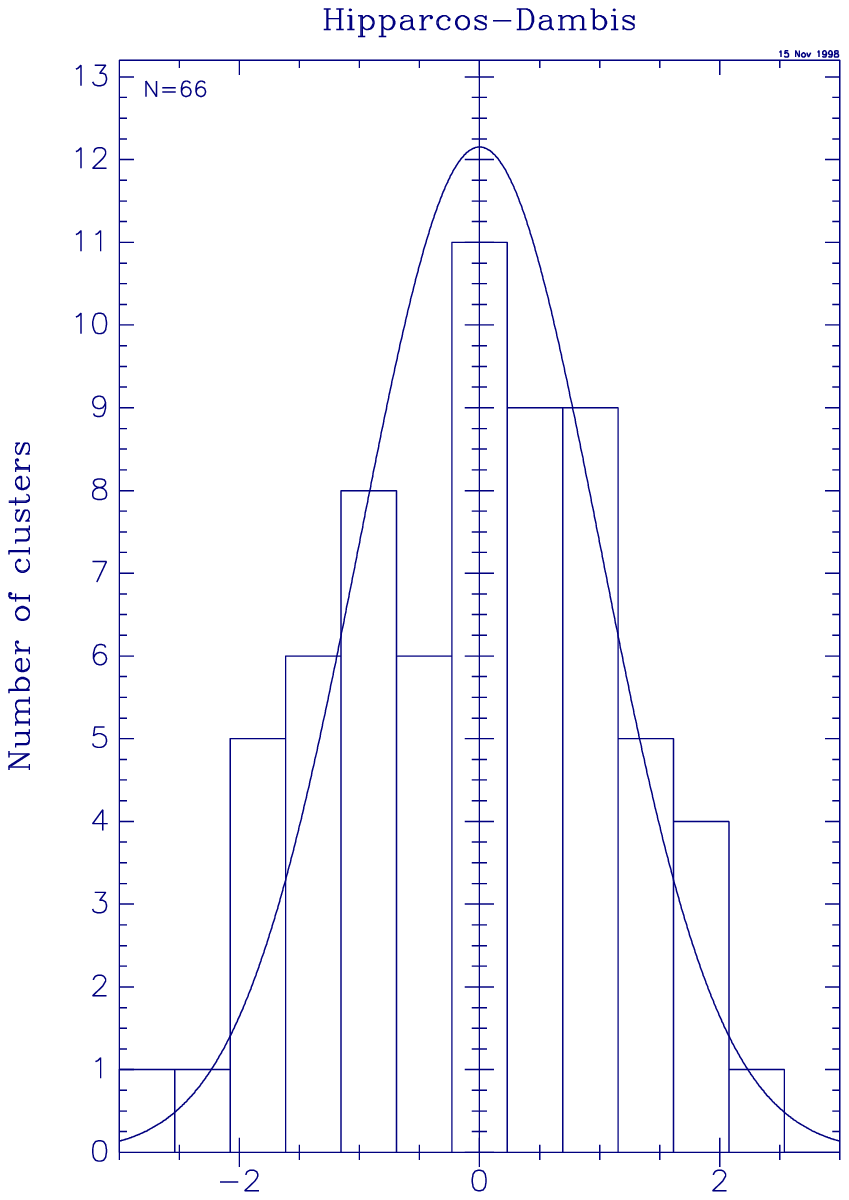}{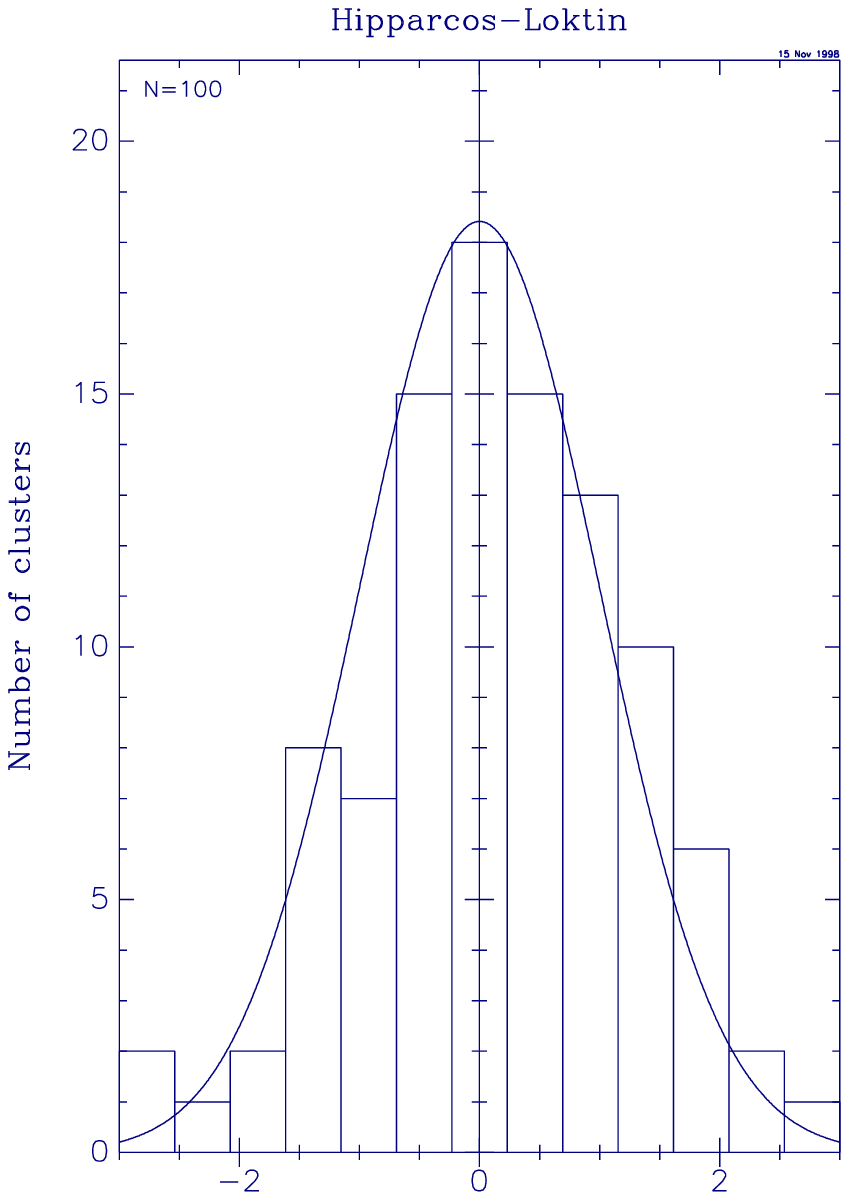}
\caption{Normalised difference between mean Hipparcos parallaxes of distant 
clusters, and the parallaxes from Dambis (1998, left) or Loktin \& Matkin 
(1994, right). A Gaussian (0,1) is superimposed.}\label{histoamas}
\end{figure}

The normalised differences between the mean Hipparcos parallaxes, taking
into account the angular correlations, and the photometric parallaxes are
shown Figure~\ref{histoamas}. It appears that the small-scale systematic
errors are 0 on the average, and the unit-weight about 1.15.  If the
Loktin \& Matkin distance moduli are corrected, taking into account the
new Hyades distance modulus (3.33 instead of 3.42), the zero-point, the
unit weight (1.17) and the asymmetry are reduced.  Since the cluster
memberships have not been thoroughly investigated, the 15\%
underestimation of the formal error on Hipparcos mean parallaxes seems to
be an upper limit. 

%
%

\begin{figure}
\plotone{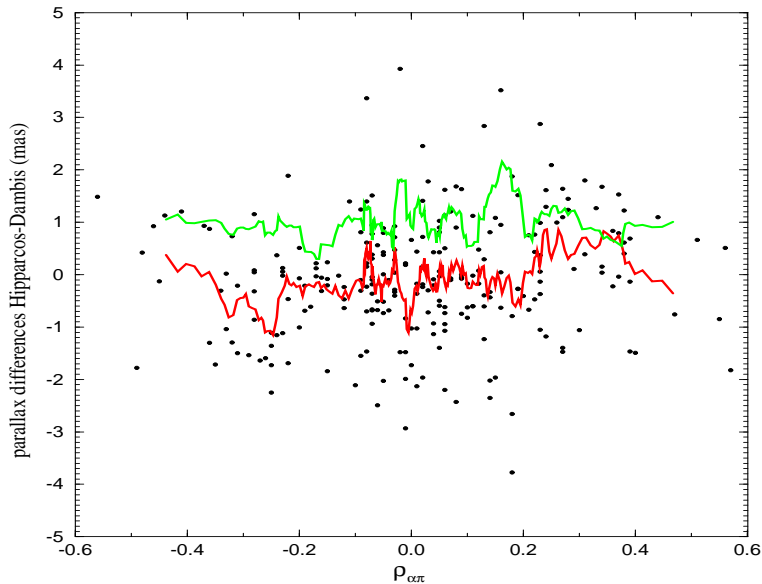}
\caption{Errors on Hipparcos parallaxes for distant clusters vs the 
correlation coefficient between right ascension and declination. The
running average and standard deviation over 10 stars is 
superimposed.}\label{rhodamb}
\end{figure}

Pinsonneault {\etal} (1998) suggested that a systematic error existed in
the mean Hipparcos parallax of the Pleiades, due to the correlations
between the right ascension and parallax, $\rho_{\alpha*\pi}$.  For each
star of the distant clusters, the difference between Hipparcos and Dambis
parallaxes is plotted Figure~\ref{rhodamb} as a function of
$\rho_{\alpha*\pi}$.  There are significant differences, due to some
clusters (NGC 6231 has a $\rho_{\alpha*\pi}\approx -0.25$), but not a
linear trend. 

%
%
From these graphs, a 1 mas systematic error for the Pleiades seems
unlikely.  We may thus assume for the following discussion that there is
no significant systematic error in the Hipparcos parallaxes, even at
small-scale.

%
\section{The effect of random parallax errors}\label{effects}
%
The astrometric elements of a star are of not much interest in themselves
for an astrophysical purpose.  Instead, the quantities of interest are the
distance, the absolute magnitude, the radius, the age or the spatial
velocity.  Given an observed parallax and proper motion, with its
associated errors, unbiased estimates of these quantities are not easy to
obtain.  For instance, it has been shown by Lutz-Kelker (1973) that a
sample selection based on the observed parallax relative error would
introduce a bias on the mean absolute magnitude.  In fact Lutz-Kelker
considered that the bias occurs at each value of the parallax,
but we will focus on sample selection only.  This bias is due to: 
\begin{itemize}
  \item the non-linear relationship between absolute magnitude (or 
  distance, etc.) and parallax, 
  \item the truncation based on the observed parallax, the true 
  parallax distribution not being uniform.
\end{itemize}
\indent These two points are discussed below, and the influence of parallax errors
is shown through the use of examples from the literature.  It should be
noted that what is true for the absolute magnitude is equally true for the
other mentioned quantities.  Although obvious, it is worth remarking that
${\sigma\over{\piH}}\propto{1\over{\piH}}$ so that the ``observed''
relative error suffers a bias, high dispersion and skewness proportional
to those of the ``observed'' distance.  The so-called Lutz-Kelker bias
occurs because the random error present in the ``observed'' relative error
is correlated with the error on the ``observed'' absolute magnitude. 

%
\subsection {Bias from non-linearity}
%
Starting from a symmetric error law for the parallaxes, the error law on
derived quantities such as distance or absolute magnitude looses this
property.  Due to their non-linearity with respect to parallax, a bias is
expected, and this is amplified by the fact that the corresponding
estimates are not defined when the observed parallax is 0 or negative,
leading to a rejection of such data. 

Given the true parallax $\pi$, and assuming a Gaussian law for the error
on the observed parallax, $\piH \leadsto {\cal N}(\pi,\sigma)$, the
expected bias of the ``observed'' distance $\rH={1\over{\piH}}$ in absence
of any truncation is
\begin{eqnarray}
E[\rH|\pi] -{1\over\pi} &=& {1\over\pi}{1\over\sqrt{2\Pi}}
\int_{-\infty}^{+\infty}\left({1\over{1+u\spipi}}-1\right)
e^{-{{u^2}\over{2}}} du \label{biais}
\end{eqnarray}
and the bias for the ``observed'' absolute magnitude 
$\MH = m + 5\log(\piH) + 5 - A$ is
\begin{eqnarray}
E[\MH|\pi] - M &=& {5\over{\sqrt{2\Pi}}}\int_{-\infty}^{+\infty}
\log(1+u\spipi)e^{-{{u^2}\over{2}}} du
\end{eqnarray}
\indent Apart from the fact that these integrals are not always defined, in both
cases a bias will be present when $\spipi$ is not negligible.  Assuming no
truncation on negative or null parallaxes, and for small relative errors,
the biases may be approximated by, respectively, $B(\rH)\approx
{1\over\pi}(\spipi)^2$ and $B(\MH)\approx -1.09(\spipi)^2$, negligible
for relative errors smaller than, say, 10\%.  This bias is due to the
asymmetry of the error distribution for $\rH$ and $\MH$, and is what would
still be present if an average of these quantities is computed;  other
statistics (based on the mode or the median) would possibly give a result
closer to the true value.

For higher relative errors, the biases and variances are depicted 
as a function of the true relative error
in Brown {\etal} (1997), Figure~1 and 2 for the distance and the absolute
magnitude respectively.

%
\subsection {Bias from truncation on observed data}
%
Whereas the bias due to the non-linearity would systematically happen (but
with a limited effect for small relative errors), the bias due to the
truncation on the ``observed'' relative error, which is the major effect,
could be avoided\ldots if no truncation was done. 

An important part of studies in the recent literature based on Hipparcos
data have used a truncation procedure, usually based on the relative
parallax error and sometimes rejecting only the negative parallaxes. In
the hope of selecting only the most precise absolute magnitudes, not only
their mean is biased, due to the Lutz-Kelker effect, but moreover the
obtained precision on this mean is worse.

A simple -- though extreme -- simulation may help to understand this fact. 
We have randomly drawn magnitude-limited samples of 1000 stars (e.g. 
RR-Lyr\ae) of constant absolute magnitude=1$^m$ with an uniform spatial
distribution\footnote{notice that in this example the samples are not
affected by Malmquist bias, even if they are limited in apparent
magnitude, because no intrinsic dispersion was introduced on the absolute
magnitudes -- see Sec. \ref{Malmquist} --. Thus, any bias will come from
non-linearity and parallax truncation}. These large samples contain on the
average only 10 stars with ``observed'' relative error better than 30\%,
and the best weighted mean of the corresponding ``observed'' absolute
magnitudes is $1.28\pm 0.28$, whereas using all stars and the estimate
discussed in section~\ref{asymp}, the mean absolute magnitude found is
$1.00\pm0.08$.  If an unweighted mean had instead been used for the
truncated sample, the bias would have reached 0.8 magnitudes.

In this example, the truncation on ``observed'' relative error gives a
30\% systematic error, of the same amount as the mean error, which is
itself 3 times greater than what would be obtained without truncation. 
The truncation thus appears as a perverse, and successful, way to obtain
both biased and unprecise results.

Although the Lutz-Kelker effect is widely known, there seems however to be
some confusion about its origin.  Since we know that the parallaxes are
individually unbiased, the average value for a {\em random} sample of
observed parallaxes will be the same as the average value of the
underlying true parallaxes, with no bias
\begin{eqnarray}
 E[\piH]&=&\int_{-\infty}^{+\infty} \piH f(\piH) d\piH =
 \int_{-\infty}^{+\infty} \piH  \int_0^{+\infty} f(\piH|\pi) f(\pi)
 d\pi d\piH\nonumber\\
 &=&\int_0^{+\infty} \int_{-\infty}^{+\infty} \piH f(\piH|\pi) d\piH 
 f(\pi) d\pi = \int_0^{+\infty} \pi f(\pi) d\pi = E[\pi]\nonumber
\end{eqnarray}
\indent On the contrary, if a truncation is done on the observed parallax
distribution (integration of $\piH$ from some limit $\pi_{-}$ in the
previous equations), there will be a bias.  Its value will depend on the
parallax distribution: for a classical magnitude-limited
sample, the measured parallax will be either too
large or too small depending on whether the truncation is done on one side
or another of the mode of the parallax distribution, as may be
deduced from $E[\pi|\piH]$ in Equation~\ref{dyson}. It must however be 
pointed out that $E[\pi|\piH]$ is {\em not} the true parallax, but an 
estimate with also a dispersion.

Thus, a sentence like ``{\it this statistical effect causes measured
parallaxes to be too large}'' (Oudmaijer et al., 1997) is likely to
mislead the reader. In fact, for one given star, few can be said when no
other information than the observed parallax is available.  Let us
consider for instance a star with observed parallax of, say, 3 mas,
which belongs to two different samples (e.g. with different limiting
magnitude), the modes of the distributions of two samples being respectively at
e.g. 2 mas and 4 mas: will the observed parallax expected to be too small or
too large? 

%
\subsection {Examples from the literature}\label{literature}
%
Since the publication of the Hipparcos Catalogue, there have been numerous
papers inferring from samples of Hipparcos stars the properties of some
populations, or comparing the new data with external data. In some cases,
the effect of random errors may be misleading, and this is mainly due to
the existing correlations between ``observed'' parallax relative errors,
``observed'' absolute magnitudes and ``observed'' distance.

A first example is taken from Tsujimoto {\etal} (1997), where the absolute
magnitudes of RR-Lyr{\ae} are calibrated.  Although the authors follow a
rigorous statistical approach, their Figure~2 may be misunderstood by the
unaware reader.  In this Figure, the ``observed'' absolute magnitude seems
to go fainter with increasing (true) distance, the stars with ``observed''
parallax relative errors greater than 100\% being systematically brighter.

\begin{figure}
\plotone{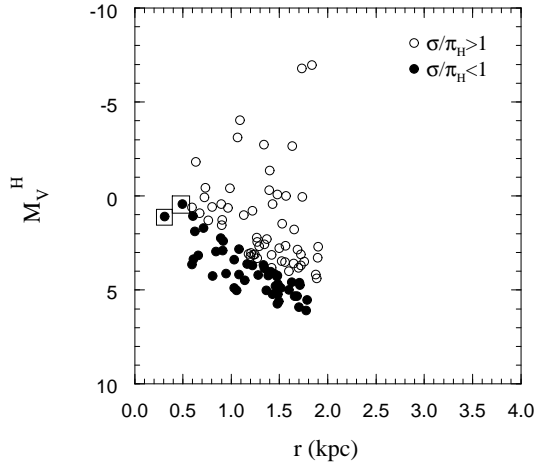}
\caption{Simulation of distant stars, showing the correlation
between ``observed'' absolute magnitude and
``observed'' parallax relative error. Only the two points indicated 
have a {\it true} relative error smaller than 1. For comparison,
see Figure~2 in Tsujimoto {\etal} (1997)} \label{moto}
\end{figure}

What could be interpreted as a systematic error in the parallax is {\em
exactly} what is expected from parallaxes with random errors - and without
systematic errors.  A simulation of 174 distant stars, assuming a constant
absolute magnitude =1, is shown Figure~\ref{moto}, excluding obviously
those with a negative parallax.  The magnitude dispersion increases with
distance (due to the increase of true parallax relative errors); the
errors bars become more and more asymmetrical, shifting some ``observed''
absolute magnitude towards the brightest end; and a positive random
parallax error implies both a fainter ``observed'' absolute magnitude and
a smaller ``observed'' parallax relative error, producing the correlation
between these two data.

%
%
A second example is taken from Oudmaijer {\etal} (1998), where the authors
discuss the Lutz-Kelker effect and apply the correction to a sample of
Cepheids.  They first compare the ``observed'' absolute magnitudes
computed using ground-based parallaxes (with a large random error) to those
computed with precise Hipparcos parallaxes as a function of the
ground-based parallax (their Figure~1, lower panel). The authors do not
seem to realise that the random errors are correlated and misinterpret
this effect as being due to a ``{\it completeness effect in the data}''. Then,
from a sample of 220 stars, only 26 stars are selected according to the
``observed'' parallax relative error. The difference between ``observed''
and true absolute magnitudes is plotted in their Figure~4. As expected,
the same effect is present, and this is not due to missing faint stars: a
volume-limited simulation will exactly reproduce the correlation effect. 

Their result in itself will not be further discussed here. As the authors
quote, Koen (1992) showed that for $\spipiH=0.175$, the 90\% confidence
interval of the Lutz-Kelker correction could span over more than 1.77
magnitudes!  Since almost all of the stars used by Oudmaijer {\etal} have
a high parallax relative error one may however wonder how their result
may be as precise as 0.02 mag. 

The two above examples illustrate the fact that the comparisons should
always be done in the plane of the measured quantities (the parallaxes),
where the errors may safely be assumed symmetrical, and not in the plane
of the derived quantities, where the effect of the random errors is not
always clear. 

In the following example, however, although the comparison is done is the
parallax plane, the effect of asymmetrical errors may be significant.  The
Figure~2 from Jahrei{\ss} et al.  (1997) shows the differences between the
Hipparcos parallaxes and those deduced from photometric CLLA parallaxes
(Carney et al., 1994) versus the CLLA parallaxes.  If there is a
systematic shift of the photometric absolute magnitudes, then it should be
seen as a slope in this graph, and the photometric parallaxes should be
corrected by a factor (1+slope).  This method could also provide an
estimate of the Hipparcos zero-point.

Although there is probably such an absolute magnitude zero-point error in
that case, it should be pointed out that there are random errors (assumed
symmetrical) in the calibrated photometric absolute magnitude, so both the
resulting asymmetrical random error of the photometric parallaxes and the
correlation between both axes may produce a similar effect (Lindegren,
1992).

\begin{figure}
\plotone{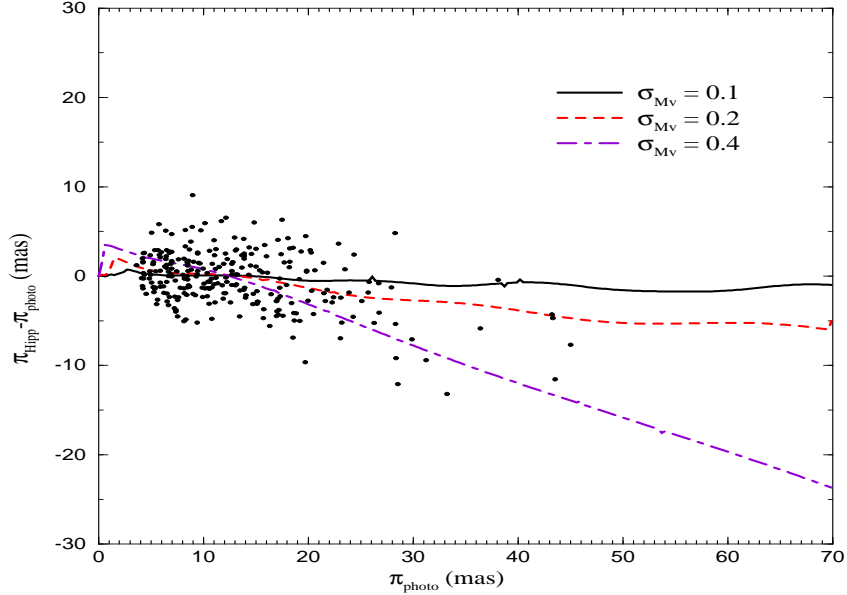}
\caption{Simulation of differences between trigonometric and 
photometric parallaxes, with the prediction as a function
of the photometric parallax error. For comparison,
see Figure~2 in Jahrei{\ss} {\etal} (1997)} \label{jahr}
\end{figure}

The way random errors may mimic a systematic error is shown
Figure~\ref{jahr}, where 275 stars have been simulated, assuming a
constant density, a linear relation between colour and absolute magnitude,
a 0.4 mag random error on absolute magnitude for the photometric estimate,
and an observed parallax computed from the true absolute magnitude.

%
%
Denoting $\piP$ the photometric parallax, the theoretical effect may 
been computed under the assumption of unbiased astrometric and 
photometric parallaxes:
\begin{equation}
E[\piH-\piP|\piP]=
{{\int_0^{+\infty}{\pi f(\piP|\pi) f(\pi) d\pi}}\over
\int_0^{+\infty}{f(\piP|\pi) f(\pi) d\pi}}-\piP
\end{equation}
\indent Assuming a Gaussian law for the distribution of the error on the 
photometric absolute magnitude, with associated variance $\sigma_M^2$
$$f(\piP|\pi) \propto 
e^{{-{1\over 2}}{{25(\log\piP-\log\pi)^2}\over{\sigma_M^2}}}$$
\noindent and assuming a magnitude-limited {\apriori} distribution for the
true parallaxes, the shape of what may be expected from only the random
errors only is shown in Figure~\ref{jahr} for different values of the
random dispersion of the photometric parallaxes.

In summary, if the random errors on the photometric absolute magnitude are
not properly taken into account in the estimation procedure, one could
wrongly deduce from such a graph both a systematic error of the absolute
magnitude and a zero-point error on the trigonometric parallaxes.  One can
not however infer from this statement that there is no systematic error in
the CLLA absolute magnitudes.

%
\section{Malmquist bias}\label{Malmquist}
%
Any finite sample of stars is, by definition, limited in apparent magnitude. In
some cases this limit can be ignored if there is another more constraining
truncation, like a limit in $\sigma\over{\piH}$. However, if one
wants to avoid as much as possible to introduce any censorship when
constructing a sample of stars, one would be left with at least an
apparent magnitude limit. It it thus important to understand the effects
that such a truncation can have on the estimation of astrophysical
quantities. 

\begin{figure}
\plotone{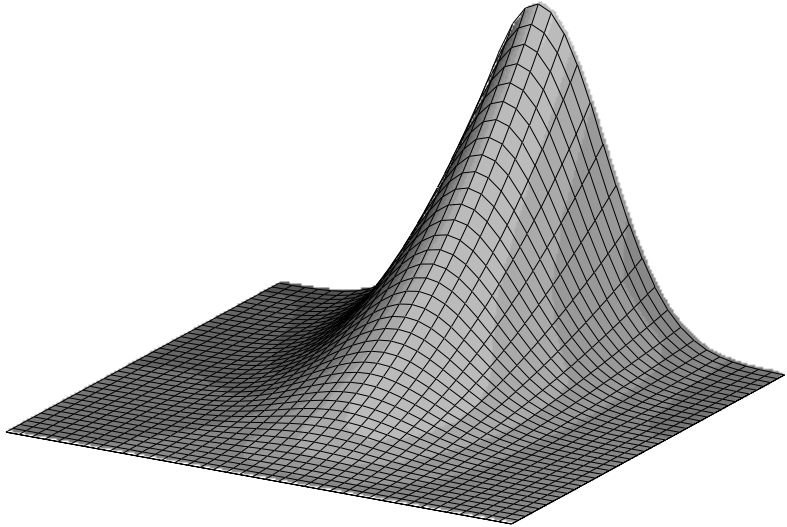}
\caption{Joint $(M,r)$ distribution for a base population with a
         Gaussian distribution in M and an homogeneous spatial
         distribution. The figure has been truncated in $r$ for
         illustration purposes.}\label{joint_1}
\end{figure}

\begin{figure}
\plotone{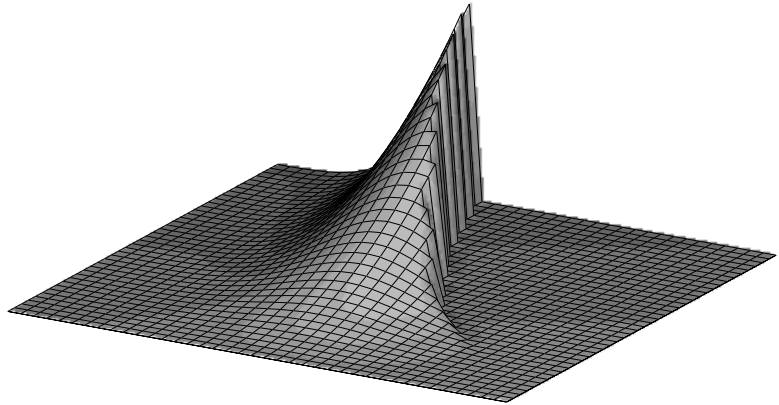}
\caption{Joint $(M,r)$ distribution for a sample with a
         Gaussian distribution in M, an homogeneous spatial
         distribution and a truncation in apparent magnitude. The figure
         has also been truncated in $r$.}\label{joint_2}
\end{figure}

The simplest case of an apparent magnitude truncation is the case of a
sample with a clean apparent magnitude limit. This case was first studied
by Malmquist (1936) under some restrictive hypothesis: 
\begin{itemize}
 \item A Gaussian distribution of the individual absolute magnitudes: 
       $M \leadsto {\cal N}(M_0,\sigma_M)$
  \item A uniform spatial distribution (space density as $r^{2}$).
\end{itemize}
\indent Under these two hypothesis the joint distribution of absolute
magnitudes and distances of the base population has the shape depicted in
Figure~\ref{joint_1}. However, when the apparent magnitude limit $m \leq
m_{\mathrm{lim}}$ is introduced this joint distribution is drastically changed, 
as depicted in Figure~\ref{joint_2}. 

%
%
%
%
While the mean absolute magnitude of the base population is $M_0$, the mean
absolute magnitude of the truncated sample $<M>$ differs from this value,
it is biased. Thus, if one uses such a sample to estimate the absolute
magnitude of the base population, even if the use of the trigonometric
parallaxes is correct the value obtained will be biased. 

\begin{figure}
\plotone{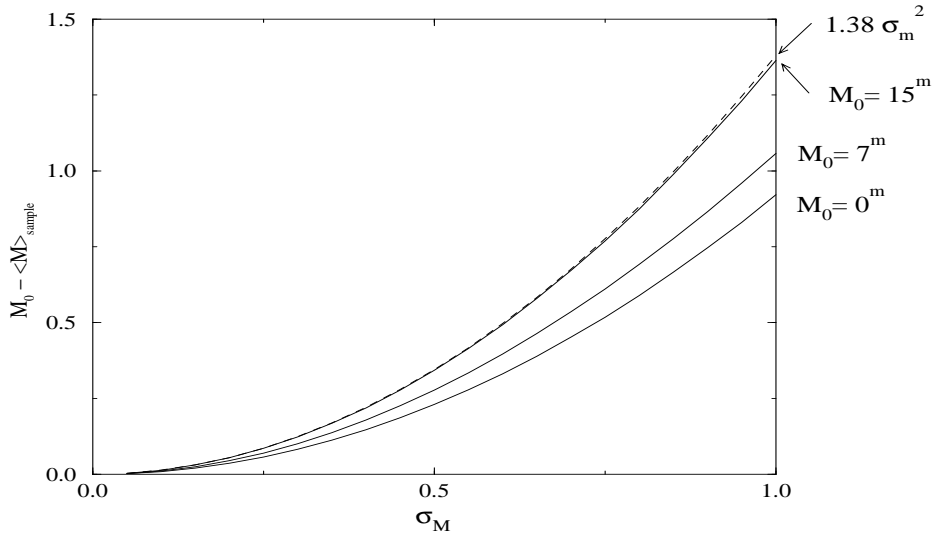}
\caption{Malmquist bias in the case of a Gaussian
         distribution of absolute magnitudes, 
         an exponential disk with $Z_0=200$ pc for the spatial distribution 
         and an apparent
         magnitude limit $m_{\mathrm{lim}}=15^m$. The classical Malmquist
         correction (dashed line) is given for comparison} \label{malmquist_Z}
\end{figure}

This bias in the mean absolute magnitude of a sample due to apparent
magnitude truncation is known as the Malmquist bias. Malmquist (1936)
calculated its value under the above cited hypothesis:
\begin{equation} \label{malmquist_formula}
  <M>\simeq M_{0}-1.38\: \sigma _{M}^{2}
\end{equation}
\indent There is, however, some confusion in the literature when using this
correction. As pointed above, the Malmquist correction is valid under the
two given hypothesis. If one of the two does not hold, the value of the
Malmquist bias may differ from Eq.~\ref{malmquist_formula}. For instance,
in the (rather common) case of an exponential disk spatial distribution
the value of the Malmquist bias depends on
$(\sigma_{M},m_{lim}-M_{0},Z_{0})$, where $Z_0$ is the scale height of the
exponential disk (Luri, 1993). An example is given in Figure
\ref{malmquist_Z}. 

%
%
Thus, Malmquist correction should not be blindly applied when an apparent
magnitude truncation is present. The correction may vary depending on the
absolute magnitude distribution and the spatial distribution of the base
population. Furthermore, if the apparent magnitude truncation is not
clean-cut the effect will also be different. This is where the Hipparcos 
Survey may come handy.

On the other hand, as can be seen in Figures \ref{joint_1} and
\ref{joint_2}, the mean distance of the sample is also biased with respect
to the base population. This can be important when studying the mean
distance of a cluster, for instance.

Finally, a further warning. All the discussion in this section has been
centred in the case of a sample truncated {\em only} in apparent
magnitude. In the case of combined truncations the joint effect should be
analysed and taken into account. For instance, as pointed out above, a
stringent truncation in $\sigma\over{\piH}$ (e.g. 10\%) may eliminate the
effects of the apparent magnitude truncation, but that may not be the case
for a less stringent truncation (e.g. 100\%). 

%
\section{Which distance and absolute magnitude from parallax?}
%
Since Dyson-Eddington (1926), who corrected the observed parallax distribution
in order to get the true absolute magnitude distribution, several methods have
been used in order to get unbiased estimates of absolute magnitudes or
distances. A first approach uses  either a transformation of these quantities,
or a correction of the biases.  Another approach uses all stars in order to
give a smaller bias.  Finally a parametrical approach together with
supplementary information fits a model to the observed quantities, taking
explicitly into account the selection biases.  The methods using a galaxy model
and simulations (e.g.  Bahcall \& Soneira, 1980, or Robin \& Cr\'ez\'e, 1986)
pertain in some sense to this latter approach but will not be discussed here.
 
%
\subsection{Transformation of the distance error law}
%
Recently, Smith \& Eichhorn (1997) have tackled the problem of distances
derived from trigonometric parallaxes.  Assuming Gaussian errors for the
parallaxes, they demonstrate the presence of bias on the ``observed''
distance and the fact that its variance can be infinite.  In this case it
would be useless to do a bias correction. Moreover the bias depends on the
true parallax relative error, which is unknown.  They propose two
different methods, using either a transformation based on the observed
parallax and its formal error, rendering a positive parallax, or a
weighting of these parallaxes, eliminating the zero parallaxes.  Each
method has advantages and disadvantages depending on whether the bias 
or the variance of the resulting estimate is considered. 

The other problem of the ``observed'' distance being its asymmetrical
error law, Kovalevsky (1998) has proposed a transformation which would
give a gaussianized distance error law for small (true) parallax relative
errors. 

It is however important to keep in mind the right use of these corrected
distances.  For instance, let us assume that we have to compare the
distances deduced from Hipparcos parallaxes to the distances deduced from
ground-based parallaxes, in order to test if there is a systematic effect
in one of the data sets.  Whereas the correct comparison would be in the plane
of parallaxes, one perverse way to do it would be to compute for the two
sets the ``observed'' distance, then to apply one of the above correction,
and finally to obtain a comparison of distances where biases are unclear
and where the high variance may prevent any safe conclusion\ldots

%
\subsection{Asymptotically unbiased estimates}\label{asymp}
%
When one needs to obtain a mean parameter on a sample, such as a mean
distance, mean absolute magnitude, etc, all parallaxes may in fact be
used, instead of computing biased estimates for each star. 

Concerning distance estimation, a simple example is the mean distance of a
cluster, neglecting the cluster depth and assuming (which is not the case
for Hipparcos) that no correlation exists between individual parallaxes.
Two possible estimates, ${\left<{1\over\pi_i}\right>}$ and
${1\over{\left<\pi_i\right>}}$ would look at first sight equivalent.  From
Equation~\ref{biais}, however, the best of these two distance estimates is
obvious: in the first case, the bias will still be present in the average
since it occurs for each inverse of parallax, although it would be
difficult to know its value since it depends on the true parallax relative
error.  Whereas, in the second case, the precision of the mean parallax
will be $\sigma\over\sqrt{n}$, so that the bias on the mean will be a
factor $\approx n$ smaller.  Asymptotically, the second estimate is thus
unbiased and should be preferred over the first, since its variance is
also smaller.  Although a bias will remain, it is in general very small
compared to all the other uncertainties: typically, a cluster at 500 pc
with only 9 Hipparcos stars will have a distance bias smaller than 3\%,
whereas an average of individual distance could give a bias greater than
30\%.

Concerning the mean absolute magnitude of a star sample, asymptotically
unbiased estimates are also used at least since Roman (1952), and detailed
in Turon \& Cr\'ez\'e (1977).  This method has recently been used by Feast
\& Catchpole (1997) or van Leeuwen \& Evans (1998) using Hipparcos
intermediate astrometric data.  The method is summarised at the end of
this section. 

However, this method concerns the mean absolute magnitude, not individual
absolute magnitudes.  The question is thus how to handle some individual
stars with poor parallax relative precision.  In general, these absolute 
magnitudes are used in an H-R diagram, e.g. for age determination or luminosity
calibrations. 
 
Instead of focusing on the absolute magnitude $M_V$, let us consider the
quantity
\begin{equation}
a_V= 10^{0.2 M_V} = \pi 10^{{m_V+5}\over 5}
\end{equation}
where the apparent magnitude $m_V$ has been corrected for extinction and
the parallax is in arcsec (or $a_V=\pi 10^{0.2 m_V-2}$ with $\pi$ in mas). 
Missing a denomination for $a_V$, we will refer in what follows to
ABL (Astrometry-Based Luminosity).  The ABL, the inverse of
the square root of a flux, is much more easy to handle than the absolute
magnitude when dealing with stars with a high parallax relative errors or
even negative parallaxes (i-e when the dispersion due to parallax random
errors is much larger than the intrinsic dispersion of absolute
magnitudes). 

In a classical H-R diagram, the absolute magnitude is plotted versus 
colour; in what we call an ``astrometric'' H-R diagram, the ABL 
is plotted versus colour.  For illustration purposes, a sample 
of 1000 stars of age 10 Gy, with [Fe/H]=-1.4 and an 0.5 mag dispersion 
in absolute magnitude, has been simulated.  No variations in 
metallicity or random errors in colours have been added.

\begin{figure}
\plotone{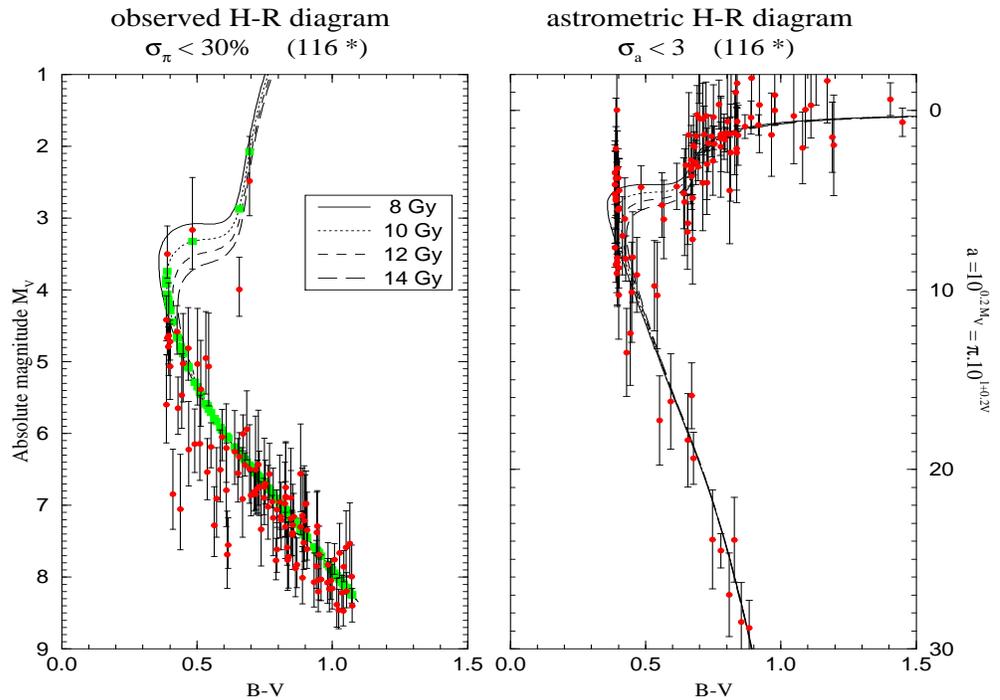}
\caption{Simulation of a sample of 10 Gy stars with [Fe/H]=-1.4 and
a 0.5 mag dispersion in absolute magnitude.
See text for legend.} \label{astrohr}
\end{figure}

The classical H-R diagram for all stars with a 30\% truncation on parallax
relative error is represented on the left of Figure~\ref{astrohr} (116
stars).  The so-called Lutz-Kelker effect appears clearly, showing the
trend to get stars below the reference line, the true position of the
stars being indicated by squares.  Since for each star the parallax
relative error is not very large, the error bar asymmetry is not well seen. 

Using the ABL, the ``astrometric'' H-R diagram is represented on the right
of Figure~\ref{astrohr}.  For the sake of comparison, the same number of
stars has been kept; this has been obtained by using
$\sigma_{a_V}<3$.  In general, there is however no reason to reject the
other stars, their high number compensates the greater error bars. 

%
%
Consider for instance a program computing the age and metallicity for a
sample of stars through interpolations between isochrones in an H-R
diagram: the truncation effect on the parallax relative error may possibly
bias the result.  On the contrary, we could get unbiased and more precise
estimates making use of the ``astrometric'' H-R diagram.  Another application 
concerns all the luminosity calibrations, the ABL being calibrated as a 
function of photometric indices. 

The use of ABL instead of absolute magnitude has the following advantages:
\begin{itemize}
\item the error bars on $a_V$ due to parallax errors are symmetrical
\item there is no Lutz-Kelker bias
\item all stars may be used, even those with negative parallaxes
\item the higher number of stars allows a gain in precision for mean values
\end{itemize}
\indent Coming back to the simple case where a mean absolute magnitude has to 
be computed from a sample of stars, and following Jung (1971) or Turon 
\& Cr\'ez\'e (1977), the first step is to estimate the best weighted 
mean ABL for the sample
$$<a_V>={{\sum_i {a_i\over{\sigma_{a_i}^2}}}
\over{\sum_i {1\over{\sigma_{a_i}^2}}}}$$
or possibly a less precise but more robust estimate,
then an asymptotically unbiased mean absolute magnitude is obtained with
$$<M_V>=5\log<a_V>$$
\indent In the case where there is an intrinsic dispersion in absolute 
magnitude (assumed small), it has to be taken into account in the weights 
of $<a_V>$.  As indicated above, all the stars may (should) be used, 
although a selection on $\sigma_{a_V}$ may be applied.  However, since 
this is a selection on luminosity, a Malmquist-type bias should be accounted 
for.  This may be also true for the whole sample. It must be pointed out 
that a symmetrical error in apparent magnitude will become asymmetrical 
in $a_V$, thus causing a bias. However, given the good photometric 
precision of Hipparcos, the bias coming from the errors in the apparent 
magnitudes is negligible and only the errors in the extinction correction 
may constitute a problem in some cases.
 
%
\subsection{Parametrical approach}
%
The approaches described above make use only of the parallax in order to
derive the distance or absolute magnitude.  Another approach makes use of
all the available information: assuming some parametrical probability
density functions (pdf), a maximum likelihood estimation allows to find
the optimal parameters corresponding to the studied sample.  An early
application of this method may be found in Young (1971), and in a more
modern way by Ratnatunga \& Casertano (1991), Arenou {\etal} (1995) and
Luri {\etal} (1996). 

Given the observables $\obs=(\piH,l,b,m_V,\mu_\alpha,\mu_\delta,V_R)$, the
parameters $\Theta$ being the coefficients of absolute magnitude as a
function of colour, the galactic scale length and scale height, the
velocity ellipsoid, etc, are estimated by maximising
\begin{eqnarray}
h(\obs;\Theta)&=&{\int_0^{+\infty}{g(\obs|\pi;\Theta)~p(\pi) d\pi}}
\mbox{ where}\\
g(\obs|\pi;\Theta)&=&p_1(\piH|\pi;\Theta)~p_2(m|\pi;\Theta)~ 
p_3(\mu_\alpha,\mu_\delta,V_R|\pi;\Theta)~p_4(l,b|\pi;\Theta)
\end{eqnarray}
where each pdf $p_i$ takes into account a possible censorship, and are
assumed to be independent; typically $p_1$ is chosen Gaussian around the
true parallax, $p_2$ is a Gaussian law for the absolute magnitude around
the mean absolute magnitude, $p_3$ is the velocity ellipsoid, and $p_4$ is
an exponential law in the galactic plane and in $Z$. The measurement error
on apparent magnitude $m$ and extinction should be taken into account in $p_2$,
as for the proper motion $(\mu_\alpha,\mu_\delta)$ or radial velocity
$V_R$ in $p_3$. An application to classical
Cepheids is given in a paper in this volume by Luri {\etal} (1998). 

As a by-product, the distance and absolute magnitude may be estimated
through e.g.  the {\aposteriori} expectation: 
\begin{eqnarray}
 \widehat{r}&=&E[1/\pi|\obs;\Theta]=
 {{\int_0^{+\infty} {1\over\pi}g(\obs|\pi;\Theta)p(\pi)d\pi}
 \over h(\obs;\Theta)}\\
 \widehat{M}&=&E[m+5+5\log\pi|\obs;\Theta]=
 m+5+5{{\int_0^{+\infty}{\log\pi}g(\obs|\pi;\Theta)p(\pi)d\pi}
 \over h(\obs;\Theta)}
\end{eqnarray}
\indent The equations above use all the known information about one given 
star, and the parameters assumed for it, so that {\it individual} 
estimates of distance, absolute magnitude, etc, may be found, even 
e.g.  if the concerned star has a zero or negative observed parallax.  
As for all Bayesian estimations, the drawback is of course that the 
{\apriori} laws must be adequate, otherwise the final result may 
be biased.

For completeness, it must be noted that there is one special case where an
{\aposteriori} estimation may be used without any {\apriori} law for the
true parallaxes, assuming only a Gaussian error law for the random
parallax errors, and making use only of the pdf of the observed parallaxes
$f(\piH)$.  This is the expectation of the true parallax given the
observed parallax, a result found by Dyson (1926).  The precision on the
obtained estimate may be also be computed: 
\begin{eqnarray}
 \widehat{\pi}&=&E[\pi|\piH]=\piH +
 \sigH^2{{f'(\piH)}\over{f(\piH)}}\label{dyson}\\
 \sigma_{\widehat{\pi}}&=& 
 \sigH\sqrt{1 + \sigH^2\left({{f'(\piH)}\over{f(\piH)}}\right)'}\nonumber
\end{eqnarray}
which is in general smaller than $\sigH$ for unimodal parallax
distributions. A more detailed discussion on the estimation of the true
parallax distribution may be found in Lindegren (1995). 

%
\section{Conclusions}
%
The Hipparcos Catalogue illustrates the various statistical problems one
has to face when fundamental parameters have to be deduced from
trigonometric parallaxes. 

The Hipparcos errors may be considered Gaussian, at least at large scales,
with no noticeable bias.  At small-scales, the correlation effect between
measurements must be taken into account.  Although the random parallax
errors are symmetrical, with zero mean and dispersion as given by the
formal error, a few outliers are however expected, e.g.  due to binarity,
in some rare cases. 

The random errors may be misleading if improperly taken into account.  In
particular the transformation of parallaxes to distance or absolute
magnitude should be done with caution. Moreover, truncations based on the
observed parallax should be avoided: although corrections to the induced
bias exist, they have large confidence intervals.

In order to estimate distances and absolute magnitudes several methods may
be used.  Either a transformation of the observed parallaxes, the use of
asymptotically unbiased estimates, or a Bayesian approach, which takes
efficiently into account the selection biases, but which rely on
{\apriori} laws. 

Apart from its numerous astrophysical applications, one of the roles of
the Hipparcos Catalogue will be to assess the validity of these {\apriori}
pdfs.  It will also assess the ground-based trigonometric parallaxes,
which will expand our knowledge to fainter stars, until new spatial
projects such as SIM or GAIA, are launched. In all cases, however, random
measurement errors will still have to be taken into account. 

%
\acknowledgments
%
Dr Lindegren, who pointed out and described an effect similar 
to the one depicted Figure~\ref{jahr} is greatly acknowledged.  We 
also thank Dr Kovalevsky who pointed to us an error in the Jacobian of 
distance in Luri \& Arenou (1997) leading to an incorrect Figure~2,
and Dr Halbwachs who provided the spectroscopic orbital data.  
An extensive use has been made of the 
SIMBAD database, operated at CDS, Strasbourg, France, and of the Base 
Des Amas (Mermilliod, 1995).

%


\begin{references}
%
\reference Arenou, F., Lindegren, L., Froeschl\'e, M., et al., 1995, \aap, 
	304, 52
\reference Arenou, F. 1997, ESA SP-1200, vol. III, chap. 17 
\reference Arenou, F., Mignard, F., Palasi J. 1997, ESA SP-1200, vol. III, 
	chap. 20 
\reference Bahcall, J. N., Soneira, R. M. 1980, \apjs, 44, 73
\reference Brown, A.G.A., Arenou, F., van Leeuwen, F., Lindegren, L.,
	Luri, X. 1997, Venise'97 symp, ESA SP 402, 63
\reference Carney et al. 1994, \aj, 107, 2240
\reference Dambis, A. K. 1998, Astronomy Letters, in press
\reference Dyson, F., 1926, \mnras, 86, 686
\reference Eddington, A.S. 1913, \mnras, 73, 359
\reference ESA 1997, The Hipparcos Catalogue, ESA SP-1200, vol. I, sect. 2.3
\reference Feast, M.W., Catchpole R.M. 1997, \mnras, 286L1
\reference Jahrei{\ss}, H., Fuchs, B., Wielen, R. 1997, 
	Venise'97 symp, ESA SP 402, p 588
\reference Jung, J. 1971, \aap, 11, 351
\reference Kovalevsky, 1998, submitted to \aap
\reference Koen C., 1992, \mnras, 256, 65
\reference van Leeuwen, F. 1997, Venise'97 symp, ESA SP 402, 203
\reference van Leeuwen, F., Hansen Ruiz, C.S. 1997, Venise'97 symp, 
	ESA SP 402, 689
\reference van Leeuwen, F. \& Evans, D. 1998, \aaps, 130, 157
\reference Lindegren, L. 1988, In `Scientific aspects of the Input Catalogue 
	Preparation II', January 1988, Sitges, J. Torra \& C. Turon eds
\reference Lindegren, L. 1992, private communication
\reference Lindegren, L. 1995, \aap, 304, 61
\reference Lindegren, L., Froeschl\'e, M., Mignard, F. 1997, 
	ESA SP-1200, vol. III, chap. 17
\reference Loktin, A.V., \& Matkin, N.V., 1994, 
	Astron. Astrophys. Trans. 4, 153
\reference Luri, X., Mennessier, M.O., Torra, J., Figueras, F. 1993,  
	\aap, 267, 305
\reference Luri, X., Mennessier, M.O., Torra, J., Figueras, F. 1996,
        \aaps, 117, 405
\reference Luri, X., Arenou F. 1997, Venise'97 symp, ESA SP 402, p 449
\reference Luri, X. 1998, this volume
\reference Lutz, T. E., Kelker, D. H. 1973, \pasp, 85, 573
\reference Malmquist, K.G. 1936, Meddel. Stockholm Obs., 26
\reference Mermilliod, J.-C. 1995, in "Information and On-Line Data in 
	Astronomy", Kluwer Academic Press, Eds D. Egret \& M.A. Albrecht, 127 
\reference Oudmaijer, R.D., Groenewegen, M.A.T., Schrijver, H. 1998, 
	\mnras, 294, L41
\reference Pinsonneault, M.H., Stauffer, J., Soderblom, D.R., King, J.R., 
	Hanson, R.B. 1998, \apj, 504, 170
\reference Robichon, N., Arenou, F., Turon, C., Mermilliod, J.C., 
	Lebreton, Y., 1997, Venise'97 symp, ESA SP 402, p 567
\reference Robichon, N., Arenou, F., Mermilliod, J.C., Turon, C.
	1999, submitted to \aap
\reference Robin, A., Cr\'ez\'e, M. 1986, \aap, 157, 71
\reference Roman, N. 1952, \apj, 116, 122
\reference Ratnatunga, K. U., Casertano, S. 1991, \aj, 101, 1075
\reference Smith, H., Eichhorn, H. 1996, \mnras, 281, 211
\reference Tokovinin, A.A. 1992, \aap, 256, 121
\reference Tokovinin, A.A., Duquennoy, A., Halbwachs, J.-L.,
	Mayor, M. 1994, \aap, 282, 831
\reference Tsujimoto, T., Miyamoto, M., Yoshii, Y. 1997, 
	Venise'97 symp, ESA SP 402, 640
\reference Turon, C.,\& Cr\'ez\'e, M. 1977, \aap, 56, 27
\end{references}
\end{document}